\documentclass[fleqn,usenatbib]{mnras}

\usepackage{newtxtext,newtxmath}
\usepackage[T1]{fontenc}

\DeclareRobustCommand{\VAN}[3]{#2}
\let\VANthebibliography\thebibliography
\def\thebibliography{\DeclareRobustCommand{\VAN}[3]{##3}\VANthebibliography}

\usepackage{graphicx}	% Including figure files
\usepackage{amsmath}	% Advanced maths commands

\def\app#1#2{% \def used due to parameters
  \mathrel{%
    \setbox0=\hbox{$#1\sim$}%
    \setbox2=\hbox{%
      \rlap{\hbox{$#1\propto$}}%
      \lower1.1\ht0\box0%
    }%
    \raise0.25\ht2\box2%
  }%
}
\def\approxprop{\mathpalette\app\relax}

\title[AGB age and mass from the P--L diagram]{Ages and masses of asymptotic giant branch stars from the period--luminosity diagram}

\author[I. McDonald]{
Iain McDonald$^{1}$\thanks{E-mail: iain.mcdonald-2@manchester.ac.uk}
\\
$^{1}$Jodrell Bank Centre for Astrophysics, University of Manchester, Oxford Road, Manchester, M13 9PL, UK
}

\date{Accepted XXX. Received YYY; in original form ZZZ}

\pubyear{\the\year{}}

\begin{document}
\label{firstpage}
\pagerange{\pageref{firstpage}--\pageref{lastpage}}
\maketitle

\begin{abstract}
A method of determining ages and masses of asymptotic giant branch (AGB) stars between 0.8 and $\sim$6\,M$_\odot$ is demonstrated, based on comparing the star's position in the period--absolute-magnitude diagram to theoretical evolutionary models. For samples of Milky Way stars, the method provides errors (statistical and systematic, respectively) of order of $^{+29}_{-35} \pm 15$ per cent in age, $^{+14}_{-7} \pm 7$ per cent in initial mass and $^{+17}_{-11} \pm 27$ per cent in current mass. However, its applicability to individual stars depends strongly on both their position in the $P-L$ diagram and the uncertainty of that position. This method is applied to published samples of AGB stars from the \emph{Gaia}, NESS, DEATHSTAR and ATOMIUM surveys. These surveys' statistical ensembles are compared to expectations from stellar evolutionary models, finding that most AGB samples are biased towards stars of younger ages and higher masses. An average mass for Milky Way AGB stars is found to be $\sim$1.1\,M$_\odot$, while mass returned to the interstellar medium by AGB stars typically comes from $\sim$1.2\,M$_\odot$ stars with mass-loss rates of order $2-3 \times 10^{-6}$\,M$_\odot$\,yr$^{-1}$.
\end{abstract}


%%%%%%%%%%%%%%%%%%%%%%%%%%%%%%%%%%%%%%%%%%%%%%%%%%

%%%%%%%%%%%%%%%%% BODY OF PAPER %%%%%%%%%%%%%%%%%%

\section{Introduction}

Asymptotic giant branch (AGB) stars represent the last nuclear-burning phase of low- and intermediate-mass stars. Several critical evolutionary processes occur during this phase, with timing and extent depending on the mass of the star. Some of these processes also occur in the more massive super-AGB stars and red supergiants (RSGs) before any truncation of their evolution in a supernova \citep[e.g.][]{Hoefner2022}. However, investigation of these processes using well-studied, nearby stars is restricted by the difficulty of measuring the AGB stars' ages and masses. Existing processes to measure masses of AGB stars rely heavily on the Hertzsprung--Russell diagram, using the core-mass--luminosity relation \citep{Blocker93,Casewell09} to measure current mass and atmospheric opacity to determine a combination of stellar envelope mass \citep[e.g.]{Kervella16,McDonald16,Shetye19,Shetye25}. However, stellar core mass only provides a limit to stellar mass through the initial--final mass function \citep{Kalirai08} while the envelope mass is strongly degenerate with metallicity, making accurate determination of mass and age essentially impossible.

The inert core of an AGB star, which will ultimately form the white dwarf, is surrounded by helium- and hydrogen-burning shells. Thermal pulses (the periodic ignition of the helium-burning shell) lead to mixing episodes whereby nuclear-enriched material can be brought to the stellar surface: a process known as third dredge-up \citep[e.g.][]{Herwig05}. This dredges up carbon-rich material, meaning the initially oxygen-rich star (spectral type M; C/O\,$\sim$\,0.4) can change into an S-star and ultimately a carbon star (C-type; C/O\,$>$\,1). In stars of more than a few solar masses, hot bottom burning favours burning of carbon to nitrogen, preventing carbon stars from forming, but still dredging up nuclear-processed material. This dredge-up process makes AGB stars an important source of many elements, particularly carbon and $s$-process elements  \citep{Kobayashi2020}. However, the efficiency of the third-dredge-up process is a major uncertainty in stellar evolution modelling.

At around the same time, layers near the surface of the star start to become unstable to pulsation. Pulsations are thought to be driven stochastically by convection cells, which excite changes in hydrogen opacity via the $\kappa$ mechanism, leading to the star occupying one of a number of pulsation sequences that run diagonally in the period--luminosity ($P-L$) diagram \citep{Wood2000,Wood2015}. These sequences, labelled on Figure \ref{fig:pl}, represent the fundamental, first-overtone and second-overtone pulsation modes. The split between sequences $B$ and $C^\prime$ appears to be triggered by the onset of the poorly understood long-secondary-period (LSP) pulsation sequence ($D$) \citep{Trabucchi2017} and corresponds to a strengthening of the pulsation amplitudes \citep{McDonald2019}.

These pulsations levitate the outer, molecule-rich layers of the star. Once material reaches $\sim$1 au ($\sim$2 stellar radii; R$_\ast$), it can cool enough to condense into either silicate- or carbon-rich dust, depending on whether the star is an M-type star or a carbon star. Radiation pressure on dust at these altitudes can exceed stellar gravity, meaning dust can be accelerated from the star. The density at this point is also high enough for strong collisional coupling between gas and dust, meaning the entrained gas is forced from the star \citep{HofnerOlofsson18}. This mass loss continues until the envelope of the star is evaporated, leaving only the inert core and (perhaps) a planetary nebula. AGB stars are therefore the dominant enrichment source for the interstellar medium in old populations with quiescent star formation, including the Milky Way \citep[e.g.][]{KarakasLattanzio14}.

Modelling these complex, dynamic stars has proved very difficult. The mass-loss rate generally exceeds the nuclear-burning rate in AGB stars, thus mass loss controls the AGB star's evolution. However, the mass-loss rate appears in turn controlled by the stellar pulsations \citep{VassiliadisWood93,McDonald2016,McDonald18}. Parameters such as mass-loss rates therefore have to be included through largely empirical models, but to be correctly modelled we need to understand the pulsations behind them. These pulsations have also proved difficult to model. Only recently have \citet{Trabucchi2021} solved this problem, by incorporating non-linear pulsation models into the  Padova {\sc parsec} evolutionary models \citep{Bressan12}. This finally means that stellar-population models can now reproduce key observables, such as the location of the pulsation sequences in the $P-L$ diagram.

We have therefore reached a juncture where a star's location in the $P-L$ diagram can be used as a determinant of its other properties. These properties notably include the age, initial mass and current mass of AGB stars, which are otherwise very difficult to recover, especially for field stars. Previous models have attempted to include such data \citep[e.g.][]{VassiliadisWood93}, but their linear and empirical pulsation models have not had the required accuracy to make good predictions.

In this work, I use a grid of {\sc parsec} models to derive statistical weights for each point on the $P-L$ diagram (Section \ref{sec:grid}) and use it to derive the ages and masses of both typical AGB stars and selected samples from the literature (Section \ref{sec:results}), closing the paper with a discussion (Section \ref{sec:disc}) and conclusions (Section \ref{sec:conc}). While this is not the first work to attempt this method and dataset \citep{Mori25}, it is the first to include fully probabilistic calculations and present generic code that can be used by others.

\section{Generating gridded ages and masses}
\label{sec:grid}

\subsection{Setup}

\begin{figure}
    \centering
    \includegraphics[height=0.47\textwidth,angle=-90]{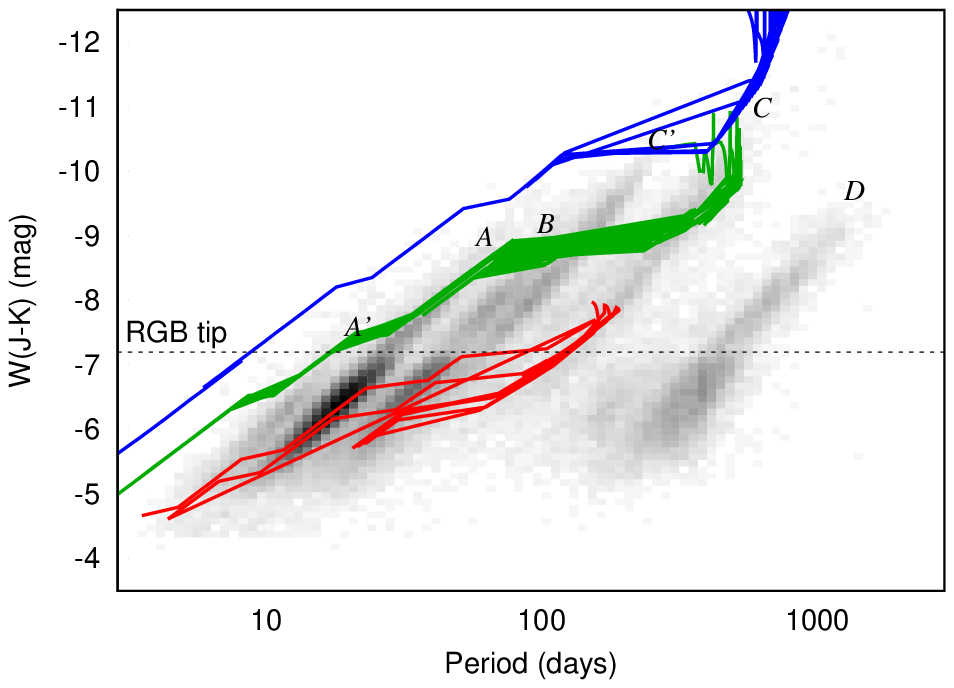}
    \caption{$P-L$ diagram for the Large Magellanic Cloud from \citet{McDonald2019}, using data from \citet{Soszynski2007} and \citet{Cutri2013}. Evolutionary tracks from the Padova models \citep{Bressan12,Trabucchi2021} are shown for populations of 10, 1 and 0.1\,Gyr (bottom to top) at a metallicity of $Z = 0.01$. Pulsation sequences are identified by their letters. The RGB tip is indicated by a dashed line.} 
    \label{fig:pl}
\end{figure}

\begin{figure}
    \centering
    \includegraphics[height=0.47\textwidth,angle=-90]{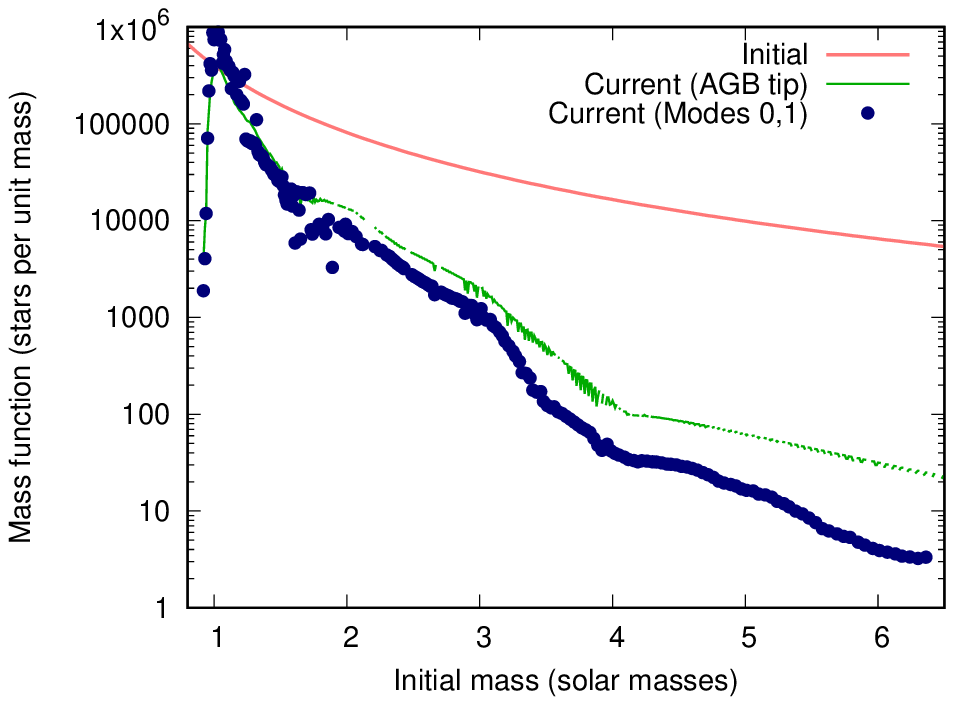}
    \caption{IMF, death function and observed AGB mass function of Galactic AGB stars, as derived from the {\sc parsec} models and \citet{Alzate2021} SFH. See text for details.} 
    \label{fig:massfn}
\end{figure}

\begin{figure}
    \centering
    \includegraphics[height=0.47\textwidth,angle=-90]{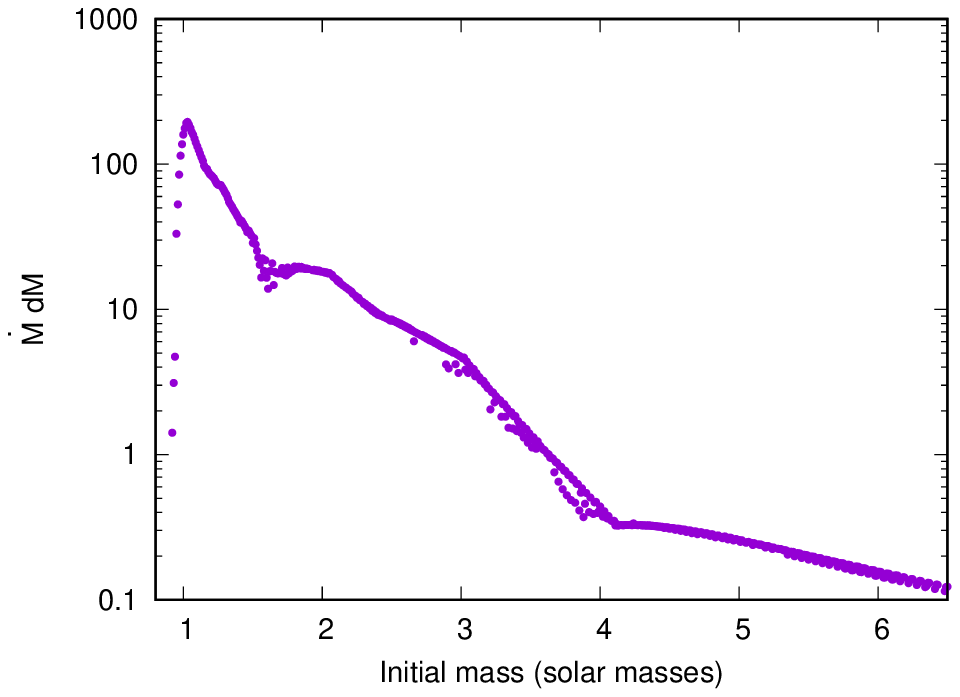}
    \caption{Relative contribution of stars of different masses to ISM enrichment (per unit time per unit volume per mass bin), based on the {\sc parsec} models and \citet{Alzate2021} SFH.} 
    \label{fig:enrichment}
\end{figure}

A set of models was generated from the Padova {\sc cmd} interface\footnote{\url{https://stev.oapd.inaf.it/cgi-bin/cmd_3.8}} with the default settings, including the {\sc parsec} v1.2S evolutionary tracks \citep{Bressan12}, the {\sc colibri S\_37} AGB tracks \citep{Pastorelli2020}, including outputs in the 2MASS $JHK_{\rm s}$ filters, using the periods and dominant pulsation regime from \citet{Trabucchi2021}. The default of ten resolution elements in the thermal pulse cycle was increased to 30 to provide greater precision and coverage of the thermally pulsating (TP-) AGB phase and thus better populate the $P-L$ diagram.

A range of isochrones was generated between ages (in years) of $\log(t) = 7.00$ and $10.13$ with a timestep of $\Delta \log(t) = 0.01$ dex. A solar metallicity of $Z = 0.0152$ was assumed and an initial set of isochrones generated at this metallicity: we return to the problem of metallicity ranges below. The resulting isochrones list not only the model star's current and initial masses, but the stellar pulsation period for modes 0 through 3 ($C$ through $A$), including which mode is dominant. They also include both the model's bolometric luminosity and photometry in the near-infrared filters\footnote{The $JHK_{\rm s}$ filters are normally used as a luminosity proxy, and often fit the $P-L$ diagram better than bolometric luminosity, though longer-wavelength filters around 3--5\,$\mu$m are often also used \citep[e.g.][]{Boyer15}. In principle, this method can be used with any filter or filter combination, though preservation of the sequences is best in the near-infrared, as it is least affected by dust absorption or emission.}.

We now need to convert these isochrones into a probabilistic frequency on the $P-L$ diagram. We can assign each point on each isochrone a weight, $w_{\rm ip}$, which defines the probability of finding a random pulsating AGB star near that grid point, given by
\begin{equation}
    w_{\rm ip} = w_{\rm IMF}(M_{\rm init}) w_{\rm SFH}(t,z) \Delta t_{\rm i} \Delta t_{\rm p} .
\label{eq:weights}
\end{equation}
Here, $w_{\rm IMF}$ is the weight applied by the initial mass function (IMF) and depends only on $M_{\rm init}$. The integrated \citep{Kroupa2001} IMF is given as part of the Padova isochrones, and differences between isochrone points can be used to determine $w_{\rm IMF}$.

The next factor, $w_{\rm SFH}$, is the weight applied by the star-formation history (SFH) of the stellar population. To apply this method to our Galaxy, we use Table 3 of \citet{Alzate2021}, which uses \emph{Gaia} stars within 100\,pc of the Sun. We generate sets of isochrones at their prescribed metallicities of $Z = 0.010,\ 0.014,\ 0.017$ and $0.030$. Linear interpolation from their irregular time grid onto our logarithmic grid was performed, preserving the relative star-formation rate at each time and at each metallicity.

The final two factors in Equation \ref{eq:weights} respectively describe the spacing in time between neighbouring isochrones and between neighbouring points on the same isochrone. For $\Delta t_{\rm i} \ll t_{\rm i}$, we can closely approximate the latter using $\Delta M_{\rm p}$, the difference in mass between neighbouring points, as $\Delta t_{\rm p} = (\Delta M_{\rm p}/\Delta M_{\rm i}) \Delta t_{\rm i}$.

For a given star, we have an observed period, $P_{\rm obs}$, with some error, $\delta P_{\rm obs}$. We also have a luminosity proxy. This can be a simple magnitude (e.g., $K_{\rm s} \pm \delta K_{\rm s}$) and distance modulus, or something more complex, such as a Wesenheit ($J-K_{\rm s}$) index \citep{Madore1982}, $W_{\rm JK} \pm \delta W_{\rm JK}$ (again, plus distance modulus), or a bolometric luminosity. Generically, we can refer to any of these as a luminosity proxy, $L_{\rm obs} \pm \delta L_{\rm obs}$ in magnitudes.

Each point on each isochrone has a dominant period ($P_{\rm ip}$) and luminosity ($L_{\rm ip}$). We can then take our variable of interest (e.g., $X = M_{\rm init}, M_{\rm current}, t, ...$) and calculate the relative probability, $P(X)$, that that variable is represented by each specific isochrone point as
\begin{equation}
    P(X) = \exp(-w_{\rm pl}/2) w_{\rm ip} ,
\end{equation}
where
\begin{equation}
    w_{\rm pl} = \sqrt{\left(\frac{P_{\rm obs}-P_{\rm ip}}{\delta P_{\rm obs}}\right)^2+\left(\frac{L_{\rm obs}-L_{\rm ip}}{\delta L_{\rm obs}}\right)^2} .
\end{equation}
The absolute probability of a star being represented by a particular point on an isochrone can be generated by normalising $P(X)$ to $\int{P(X)} dX = 1$. By sorting in $X$ this array of $P(X)$, we can generate a statistical probability distribution in $X$. The cumulative distribution can then be used to identify, e.g., the median probability and the 68\% confidence interval. Similar calculations can be performed in frequency, which can be useful for dealing with asymmetric errors in period.

\subsection{Limitations}

\begin{figure}
    \centering
    \includegraphics[height=0.47\textwidth,angle=-90]{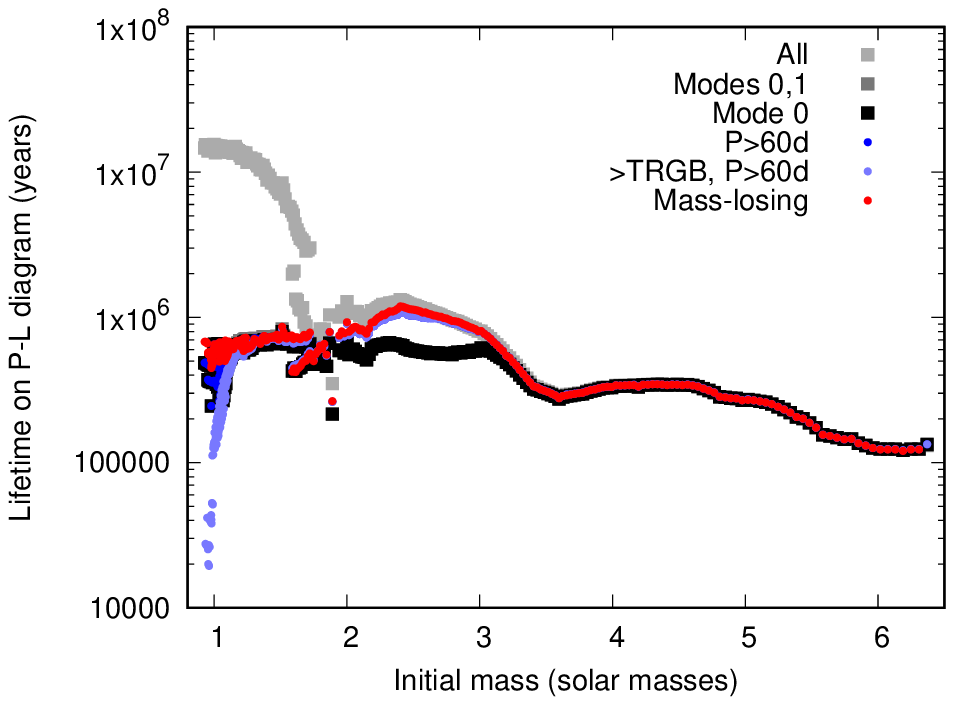}
    \includegraphics[height=0.47\textwidth,angle=-90]{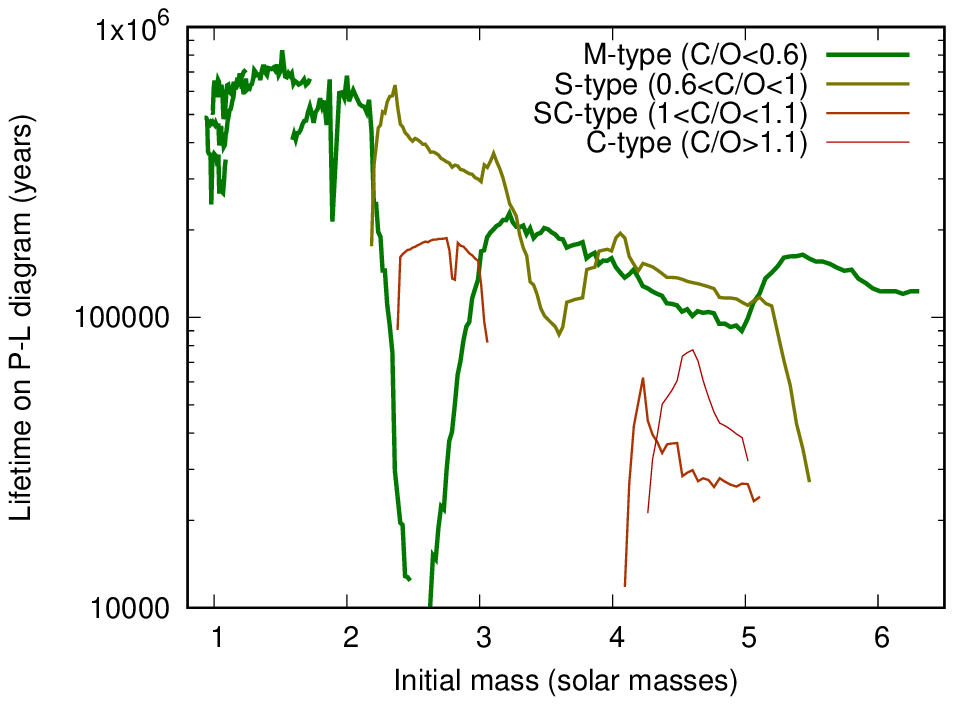}
    \includegraphics[height=0.47\textwidth,angle=-90]{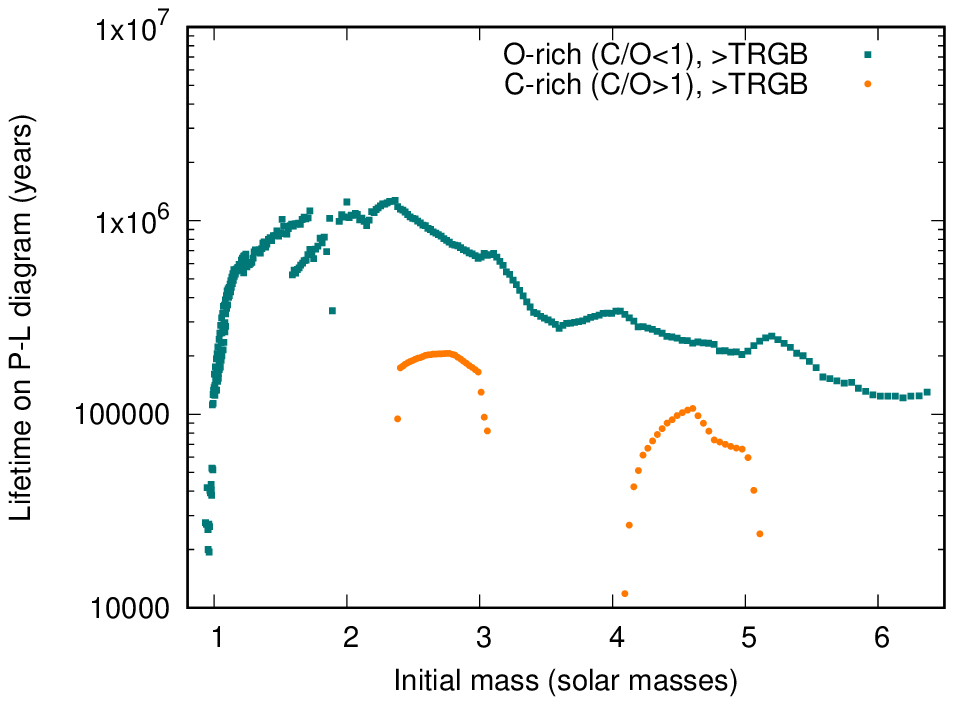}
    \caption{Time spent as pulsating evolved stars under different boundary conditions. Top panel: lifetime when limited by pulsation mode (fundamental, overtone) or period ($>$60\,days), luminosity (above the RGB tip) or strong mass-loss rate ($>$10$^{-7}$\,M$_\odot$\,yr$^{-1}$). Middle panel: stars with the above mode, period and mass-loss limits, divided into lifetimes by their C/O ratios as indicated. Bottom panel: stars above the RGB tip luminosity, split into O-rich and C-rich chemistries as indicated. Based on the {\sc parsec} models and \citet{Alzate2021} SFH: note the artificial gap in carbon-star production around 3--4\,M$_\odot$ due to the coarse metallicity sampling of the latter. See text for full details.}
    \label{fig:lifetime}
\end{figure}

There is no one-to-one mapping of a point on the $P-L$ diagram and the age or mass of a star: multiple evolutionary tracks cross through the same point. It therefore needs stressing that the ages and masses obtained for AGB stars using this method are statistical estimates, not direct measurements. Derived ages and masses are also reliant on the input models and any deviations of these from reality will be propagated into the resulting ages and masses.

Errors in these models may include:
\begin{itemize}
    \item Errors in the shape of the underlying \citet{Kroupa2001} IMF.
    \item Errors in the underlying SFH. For this paper, it is noteworthy that only broad categorisations of metallicity are included and the most-metal-poor populations do not form part of the input dataset. It is particularly important to change the star-formation history to one appropriate to the population in question.
    \item Statistical effects caused by the evolution of stars. It is assumed here that the death function of a stellar population is simply the IMF truncated by normal single-star evolution. Unmodelled binary interaction may cause deviations from this. Among other effects, these include the formation of extrinsic carbon stars, which are not modelled here.
    \item Evolutionary and atmospheric differences caused by abundances. The Padova isochrones assume solar-scaled abundances, whereas many stellar populations have substantial differences in elemental abundances like helium or [$\alpha$/Fe]. This can affect both stars' evolutionary tracks and their atmospheric opacity, thus their pulsation properties.
    \item Errors in the Padova stellar evolution models. The AGB phase is notoriously difficult to model (see the Introduction), and errors in previous evolutionary steps can compound. Such errors can include but are not limited to: errors in the treatment of mixing, dredge-up and hot bottom burning, especially during thermal pulses; errors caused by assumptions about stellar rotation and magnetic fields; errors caused by radiative transfer in atmospheric models, especially with approximations of local thermodynamic equilibrium in strongly pulsating stars and with the treatment of circumstellar dust; errors in the estimated mass-loss rates; and errors in determining the fundamental period, especially in the non-linear regime, and determining the dominant period from the pulsation growth and decay rates. These cause associated uncertainties and errors related to the initial--final mass function and the maximum modelled mass for an AGB star. Most notably, there is an effective cap at $M_{\rm init} \approx 6$\,M$_\odot$, beyond which the AGB wind does not develop in the isochrones, since the length of the TP-AGB approaches zero. (The stars appear on the $P-L$ diagram but do not undergo their full mass loss.)
\end{itemize}
Care should also be taken with the following observational points:
\begin{itemize}
    \item The models do not account for any binary companions that may either influence the stellar evolution or the apparent luminosity of the AGB star.
    \item The models assume mass loss is spherically symmetric. In stars where dust opacity is significant (in the bands used to measure luminosity), deviations from spherical symmetry may affect the position on the $P-L$ diagram.
    \item Uncertainties in pulsation periods are often not given, and some surveys give erroneously small periods or the wrong pulsation modes if their window of observation is not long enough.
    \item Care should be taken to ensure the dominant pulsation mode is correct. Mixed-mode pulsations are not considered. Sequence D pulsations are not modelled as part of \citet{Trabucchi2021}. If a star is placed outwith the $P-L$ sequences, a null or highly uncertain result will be generated.
    \item Luminosity proxies should take into account interstellar reddening, including for Wesenheit indices.
    \item Luminosity uncertainties should account for errors caused by variability. If the uncertainties remain small, calibration errors in the surveys (including errors in filter-transmission functions and colour-corrections for red stars) may need to be taken into account.
    \item Distance uncertainties for variable stars may be underestimated. Note that stars with large parallax uncertainties may have distance moduli uncertainties that are non-Gaussian \citep[cf.,][]{LutzKelker73}. Due to their large size and surface motions, parallax uncertainties may also be under-estimated \citep[e.g.][]{Chiavassa2018,Uttenthaler26}.
    \item Errors in period and brightness may be correlated in low signal-to-noise lightcurves.
\end{itemize}

\section{Results}
\label{sec:results}

\subsection{Evolved-star mass functions and ISM return}

Not all AGB stars will pulsate in the modelled modes ({\tt Mode = 0-4}); $C$, $C^\prime$/$B$, $A$ and $A^\prime$), and AGB stars of different mass and metallicity will enter these modes at different times and progress through them at different speeds. We can therefore begin by looking at the expected population of AGB stars in the $P-L$ diagram, before comparing this expected distribution to observational samples of AGB stars.

Figure \ref{fig:massfn} shows the convolution of the IMF with the Milky Way SFH to derive the ``death function'': the expected initial mass function of the subset of stars leaving the AGB. This death function distribution differs substantially from the IMF. Partly this is due to the star-formation rate, which has declined by a factor of $\sim$30 over the last 10 Gyr, thus favouring lower-mass stars. However, it is also partly due to the scaling of main-sequence lifetime to stellar mass, which adds an \emph{additional} $M^{-1.5}$ scaling to the number of stars per unit mass leaving the AGB today. The death function of today's AGB stars therefore has an approximate scaling of $\xi(M) \approxprop M^{-2.3}M^{-1.5}M^{-1.5} \approxprop M^{-5.3}$.  The exact scaling is dependent on details of both SFH and stellar evolution, giving the green line in Figure \ref{fig:massfn} (the noisy data around 1.3 and 1.8\,M$_\odot$ are artificially scattered due to the change in assigned metallicity at this point and finite grid spacing of the isochrones).

Multiplying this death function by the total mass lost at each stellar mass (here, by invoking the initial--final mass function of \citet{Cummingham24}), we can derive the relative contribution of stars of different masses to the enrichment of the ISM (Figure \ref{fig:enrichment}). Roughly half the mass entering the ISM comes from stars of $M < 1.22$\,M$_\odot$, 90\% from stars of $M < 2.32$\,M$_\odot$, 95\% from stars of $M < 2.78$\,M$_\odot$ and 99\% from stars of $M < 3.97$\,M$_\odot$. This does not account for the small contributions from super-AGB stars, red supergiants (RSGs) or (super-)novae. It does include mass loss from stages before the AGB (notably the red giant branch (RGB)), which can dominate over AGB mass loss in the lowest-mass stars \citep[e.g.][]{McDonaldZijlstra15}.

\subsection{Current observable AGB mass function}
\label{sec:cmf}

The observed population of pulsating AGB stars also depends on the lifetime of the pulsating AGB phase. Since the mass-loss rate exceeds the nuclear-fusion rate, the lifetime is dependent on losing the star's atmospheric mass. While this should take proportionally longer for more massive stars, at higher masses, this effect is countered by the increased mass-loss rates that massive stars endure.

At this point, the semantic definition of a pulsating AGB star becomes important. Low-mass stars also undergo small-amplitude (``SARG'') pulsations while still on the red giant branch (RGB) \citep[][see stars below the RGB tip in Figure \ref{fig:pl}]{Soszynski2007}, and can therefore spend several Myr on the $P-L$ diagram. However, these stars may only have periods of a few days and amplitudes of a few mmag, so are not typically included in catalogues of AGB stars as they lack both strong pulsations and the associated strong, dusty winds. Moderately strong pulsations can start in AGB stars at comparatively low luminosities ($\sim$700--1000\,L$_\odot$; \citealt{McDonald11,McDonald17}), but such stars can be difficult to separate from RGB stars \citep[cf.,][]{Groenewegen12,McDonald16}, meaning these stars are often not included either. Furthermore, evolution across the $P-L$ diagram is initially slow while stars' mass loss is weak and their evolution remains dominated by nuclear burning inside the star, but stars quickly transition to longer periods and stronger pulsations as mass-loss rates increase.

Consequently, these fainter, weakly pulsating ``SARG'' stars represent the majority of the stars on the $P-L$ diagram and, if we are to correctly estimate distributions of AGB stars that can be compared to observed samples, we must place limitations on the model data according to our observed selection function. Common selection functions may incorporate (either specifically or implicitly) limits on one or more of the following criteria:
\begin{itemize}
    \item {\it Luminosity.} In many surveys, only stars above the luminosity of the RGB tip are included (here, we assume $L_{\rm TRGB} \approx 2500$\,L$_\odot$).
    \item {\it Pulsation period, mode or amplitude.} Searches for long-period variables (LPV) often implicitly include lower limits on pulsation amplitude. This is difficult to model accurately, but is intrinsically tied to pulsation period and mode. Typically, short-period stars do not pulsate strongly enough ($\delta V \gtrsim 0.1$\,mag) to achieve the status of LPV or semi-regular variable (SRV), with the cutoff being mass-dependent but no less than $P \approx 60$\,days \citep{McDonald2016}. Large-amplitude pulsation can also be tied to pulsation mode, with LPV/SRV stars almost universally inhabiting pulsation modes $C$ and $C^\prime$ \citep{McDonald2019}. We can therefore restrict models to \citet{Trabucchi2021}'s {\tt Mode = 0-4} to select all variable stars; {\tt Mode = 0} or {\tt 1} to approximate surveys for strong variability\footnote{{\tt Mode = 1} also includes sequence $B$, since both $B$ and $C$ are first-overtone pulsations, but this is the closest approximation we can make.}; or {\tt Mode = 0} to approximate surveys for classical Miras\footnote{However, without proper luminosity constraint, this will also include many SRV stars.}.
    \item {\it Mass-loss or dust-production rate.} Since mass loss and dust production is strongly linked to pulsation amplitude \citep{McDonald18}, selecting {\tt Mode = 0} or {\tt 1} will also serve to select stars exhibiting strong, dusty mass loss \citep{McDonald2019}. However, we can also specifically target stars that are producing dusty winds in the models by selecting, e.g., stars with mass-loss rates $\dot{M} > 10^{-7}$\,M$_\odot$\,yr$^{-1}$.
    \item {\it C/O ratio.} Finally, we can separate stars according to C/O ratio, dividing stars into O-rich and C-rich stars, which we define simply by limits in C/O ratio of $<$1 and $>$1, respectively. This is often done photometrically for star counts for C/M-star ratios, and performed irrespective of mass-loss rate, so we match this C/O limit with our luminosity limit of stars above the RGB tip. Alternatively, for spectroscopic surveys, we can further classify M-type, S-type, SC-type and C-type stars with C/O ratios of $<$0.6, 0.6--1.0, 1.0--1.1 and $>$1.1, respectively. Since this is normally linked to strongly pulsating, dust-producing stars, we match this with our limits of {\tt Mode = 0} or {\tt 1}, $P > 60$\,days and $\dot{M} > 10^{-7}$\,M$_\odot$\,yr$^{-1}$.
\end{itemize}

Care must be taken with these latter two criteria. The mass-loss rates in the {\sc parsec/colibri} evolution models in particular are based on sophisticated models and comparison against both reference populations and the initial--final mass function of stars. However, particularly in chemically extreme stars (C/O $\gg$ 1 or [Fe/H] $\ll$ 0\,dex), the mass-loss rate at any given point in time lacks both good inference from first principles or empirical calibration. Consequently, derived masses and ages may become unreliable, especially for higher mass-loss or dust-production rates.

Similarly, limits based on C/O ratio assume accurate modelling of complex nuclear and mixing processes, including hot bottom burning and dredge-up during thermal pulses, which are some of the most poorly constrained factors in AGB evolution modelling. Additionally, it requires an age--metallicity relation with a finer structure in metallicity than that applied here, in order to properly capture the differences in natal carbon abundance that restrict the formation of carbon stars at high metallicity. Finally, we remind the reader that the SFH used here is based on stars within 100\,pc of the Sun, and the methodology therefore cannot be directly applied to environments such as the Galactic Halo without first changing the SFH. Consequently, the applications of this method later in the paper are restricted to luminosity and pulsation properties, which are much better modelled.

These issues notwithstanding, Figure \ref{fig:lifetime} shows the lifetime of stars in these various phases, as reconstructed from the {\sc parsec} isochrones. The large peak generated by RGB ``SARG'' pulsators can be seen in the upper left of the top panel (grey points), which disappears when stars are restricted to those brighter than the RGB tip. Due to the comparatively short time that stars spend on sequences $B$ and $C$, the lifetime of stars as strong pulsators ({\tt Modes 0} and {\tt 1}) and fundamental-mode ($C^\prime$) pulsators ({\tt Mode 0}) are negligibly different from each other, and (again due to the rapid crossing of sequences) negligibly different from the all-modes case for higher ($M \gtrsim 3$\,M$_\odot$) masses.

Figure \ref{fig:lifetime} also shows the lifetime spent as chemically enhanced S-type, SC-type and C-rich stars. An artificial decrease between $\approx$3 and 4\,M$_\odot$ occurs here due to the step change in metallicity from $Z = 0.017$ (which populates the low-mass end) to $Z = 0.03$ (which populates the high-mass end). This step change exemplifies exemplifies the current limitations in applying this method to specific chemical classes of stars. This also affects the presence of SC-type stars, which are only seen in the models at higher masses (therefore younger populations), due to the longer time metal-rich stars spend in this phase.

Multiplying the death function in Figure \ref{fig:massfn} by the AGB lifetime (in case shown in Figure \ref{fig:massfn}, for pulsations on {\tt Mode = 0} and {\tt 1}) yields the observed AGB mass function. As with previously described functions, the observed AGB mass function shows a sharply peaked function that strongly concentrates observable AGB stars in a population near 1\,M$_\odot$.

\subsection{AGB general properties}

By integrating over the observed AGB mass function (defined with appropriate limits in observable properties), we can derive the probability of an arbitrary observed AGB star lying within those limits being of a given mass. These probabilities are shown in Table \ref{tab:typical} for the different limits listed in Section \ref{sec:cmf}, with median values, and 1$\sigma$, 2$\sigma$ and 3$\sigma$ (68, 95 and 99.5 per cent) confidence intervals. Given our applied SFH, these entries identify the expected typical properties of pulsating AGB stars within 100\,pc of the Sun.
 
Some scientific applications, such as enrichment of the ISM, will be more interested in where the mass return by stars is focussed. Evidence from metal-poor galaxies with ongoing star formation indicates that mass loss is dominated by a small fraction of ``extreme'' AGB stars \citep{Blum06,Boyer11}. To address these applications, Table \ref{tab:typicalMLR} provides the expected properties of AGB stars if they are sampled proportionally to their mass-loss rate. This indicates the typical masses of stars from which replenishment of the Galactic ISM originates. The comparatively early peak in the Galactic SFH means that the typical AGB enrichment is performed by lower-mass, less-extreme stars than in smaller Local Group galaxies.

Tables \ref{tab:typical} and \ref{tab:typicalMLR} demonstrate that most stars on the $P-L$ diagram descend from solar-mass stars. This is due to both their high frequency (Figure \ref{fig:massfn}) and the long time these stars spend on the diagram, including as both RGB stars (Figure \ref{fig:lifetime}) and faint stars below the luminosity of the RGB tip \citep[e.g.][]{McDonald17}. The tables also show that most of these are yet to lose significant mass, which reflects the accelerating rate of mass loss over the AGB lifetime.

Placing limits on either pulsation sequences or stars with measurable dust production (using $\dot{M} > 10^{-7}$\,M$_\odot$\,yr$^{-1}$ as a proxy) only serve to re-enforce this concentration to lower masses, since low-mass stars reach the fundamental mode quickly and stay on it losing modest amounts of mass for a long time.

The only significant departures in Tables \ref{tab:typical} and \ref{tab:typicalMLR} from this solar-mass hegemony are selections based on either luminosity or chemistry. Only higher-mass stars reach higher luminosities and, the larger the star, the longer they remain above any given luminosity. Selections based on C/O ratio focus the distribution towards higher-mass stars, where third dredge-up is effective, while avoiding the highest-mass stars where hot-bottom burning destroys carbon. In both these cases, the observed distribution narrows towards younger, higher-mass stars. This shift also occurs if a mass-loss-rate limits of $\dot{M} \gg 10^{-7}$\,M$_\odot$\,yr$^{-1}$ are chosen, as stars with high mass-loss rates are generally higher-mass and often carbon-rich stars. However, the reader is reminded that modelled mass-loss rates and chemical types may not be entirely accurate, due to both limitations of the models and their sampling by the above SFH.

Different sampling methods for selecting AGB stars can therefore result in wildly different populations of stars being sampled. For instance, the populations from which a sample of 100 oxygen-rich and 100 carbon-rich stars are drawn may have zero overlap (at least in metal-rich populations). Care therefore needs to be taken when concluding general facts from specific samples of AGB stars.

\begin{table}
    \centering
    \caption{Expected properties of AGB stars in a statistically unbiased Galactic $P-L$ diagram.}
    \label{tab:typical}
    \begin{tabular}{@{}l@{\quad}c@{\quad}c@{\quad}c@{\quad}c@{}}
    \hline \hline
Parameter                   & Median    & 1$\sigma$ range   & 2$\sigma$ range   & 3$\sigma$ range \\
\hline
\multicolumn{5}{c}{All stars ({\tt Mode = 0-4})}\\ %extracts-all.dat
Age (Gyr)                   & 8.13 & 4.37 -- 10.7 & 2.09 -- 12.6 & 0.56 -- 13.2 \\
$M_{\rm init}$ (M$_\odot$)  & 1.10 & 1.01 -- 1.32 & 0.96 -- 1.76 & 0.94 -- 2.89 \\
$M_{\rm curr}$ (M$_\odot$)  & 1.07 & 0.97 -- 1.30 & 0.89 -- 1.76 & 0.80 -- 2.89 \\
\multicolumn{5}{c}{Mass-losing stars ($\dot{M} > 10^{-7}$\,M$_\odot$\,yr$^{-1}$)}\\ %extracts-masslosing.dat
Age (Gyr)                   & 10.2 & 8.13 -- 11.8 & 2.63 -- 12.9 & 0.68 -- 13.2 \\
$M_{\rm init}$ (M$_\odot$)  & 1.03 & 0.99 -- 1.10 & 0.96 -- 1.60 & 0.94 -- 2.73 \\
$M_{\rm curr}$ (M$_\odot$)  & 0.83 & 0.76 -- 0.92 & 0.70 -- 1.41 & 0.62 -- 2.55 \\
\multicolumn{5}{c}{LPVs ($P > 60$\,days)}\\ %extracts-gt60days.dat
Age (Gyr)                   & 10.2 & 8.32 -- 11.8 & 2.00 -- 12.9 & 0.04 -- 13.2 \\
$M_{\rm init}$ (M$_\odot$)  & 1.03 & 0.99 -- 1.10 & 0.96 -- 1.83 & 0.94 -- 8.08 \\
$M_{\rm curr}$ (M$_\odot$)  & 0.82 & 0.75 -- 0.90 & 0.69 -- 1.68 & 0.61 -- 8.07 \\
\multicolumn{5}{c}{LPVs ($P > 60$\,days) above RGB tip ($L > 2500$\,L$_\odot$)}\\ %extracts-gt60days-trgb.dat
Age (Gyr)                   & 6.46 & 2.57 -- 10.0 & 0.55 -- 11.5 & 0.03 -- 12.9 \\
$M_{\rm init}$ (M$_\odot$)  & 1.18 & 1.03 -- 1.61 & 1.00 -- 2.92 & 0.96 -- 8.79 \\
$M_{\rm curr}$ (M$_\odot$)  & 0.92 & 0.76 -- 1.42 & 0.64 -- 2.74 & 0.59 -- 8.78 \\
\multicolumn{5}{c}{Sequences $B$, $C$ and $C^\prime$ ({\tt Mode = 0,1})}\\ %extracts-bccp.dat
Age (Gyr)                   & 10.5 & 8.51 -- 11.8 & 2.95 -- 12.9 & 0.57 -- 13.2 \\
$M_{\rm init}$ (M$_\odot$)  & 1.03 & 0.99 -- 1.09 & 0.96 -- 1.52 & 0.94 -- 2.88 \\
$M_{\rm curr}$ (M$_\odot$)  & 0.82 & 0.75 -- 0.90 & 0.69 -- 1.34 & 0.61 -- 2.58 \\
\multicolumn{5}{c}{Sequence $C^\prime$ ({\tt Mode = 0})}\\ %extracts-cp.dat
Age (Gyr)                   & 10.5 & 8.51 -- 12.0 & 2.95 -- 12.9 & 0.57 -- 13.2 \\
$M_{\rm init}$ (M$_\odot$)  & 1.03 & 0.99 -- 1.09 & 0.96 -- 1.52 & 0.94 -- 2.88 \\
$M_{\rm curr}$ (M$_\odot$)  & 0.81 & 0.75 -- 0.89 & 0.69 -- 1.33 & 0.61 -- 2.59 \\
\multicolumn{5}{c}{O-rich stars (C/O\,$<$\,1; $L > 2500$\,L$_\odot$)$^\ast$}\\ %extracts-orich.dat
Age (Gyr)                   & 1.20 & 0.89 -- 7.41 & 0.35 -- 10.5 & 0.04 -- 12.6 \\
$M_{\rm init}$ (M$_\odot$)  & 2.25 & 1.13 -- 2.49 & 1.02 -- 3.40 & 0.97 -- 8.40 \\
$M_{\rm curr}$ (M$_\odot$)  & 2.23 & 0.85 -- 2.49 & 0.69 -- 3.38 & 0.60 -- 8.39 \\
\multicolumn{5}{c}{C-rich stars (C/O\,$>$\,1; $L > 2500$\,L$_\odot$)$^\ast$}\\ %extracts-crich.dat
Age (Gyr)                   & 0.76 & 0.59 -- 0.93 & 0.50 -- 1.00 & 0.15 -- 1.02 \\
$M_{\rm init}$ (M$_\odot$)  & 2.63 & 2.46 -- 2.86 & 2.40 -- 3.01 & 2.38 -- 4.61 \\
$M_{\rm curr}$ (M$_\odot$)  & 1.82 & 1.16 -- 2.26 & 0.90 -- 2.47 & 0.81 -- 2.55 \\
\multicolumn{5}{c}{M-type stars (C/O\,$<$\,0.6; {\tt Mode = 0,1})$^\ast$}\\ %extracts-m.dat
Age (Gyr)                   & 10.5 & 8.51 -- 11.8 & 3.24 -- 12.9 & 1.35 -- 13.2 \\
$M_{\rm init}$ (M$_\odot$)  & 1.03 & 0.99 -- 1.09 & 0.96 -- 1.47 & 0.94 -- 2.17 \\
$M_{\rm curr}$ (M$_\odot$)  & 0.81 & 0.75 -- 0.90 & 0.69 -- 1.29 & 0.61 -- 2.02 \\
\multicolumn{5}{c}{S-type stars (0.6\,$<$ C/O\,$<$ 1; {\tt Mode = 0,1})$^\ast$}\\ %extracts-s.dat
Age (Gyr)                   & 0.98 & 0.68 -- 1.18 & 0.45 -- 1.29 & 0.20 -- 1.32 \\
$M_{\rm init}$ (M$_\odot$)  & 2.42 & 2.27 -- 2.73 & 2.20 -- 3.13 & 2.19 -- 4.16 \\
$M_{\rm curr}$ (M$_\odot$)  & 2.19 & 1.94 -- 2.49 & 1.00 -- 2.78 & 0.69 -- 3.48 \\
\multicolumn{5}{c}{SC-type stars (1\,$<$ C/O\,$<$ 1.1; {\tt Mode = 0,1})$^\ast$}\\ %extracts-sc.dat
Age (Gyr)                   & 0.76 & 0.59 -- 0.93 & 0.50 -- 1.00 & 0.18 -- 1.02 \\
$M_{\rm init}$ (M$_\odot$)  & 2.63 & 2.46 -- 2.86 & 2.40 -- 3.01 & 2.38 -- 4.26 \\
$M_{\rm curr}$ (M$_\odot$)  & 1.80 & 1.16 -- 2.24 & 0.89 -- 2.47 & 0.81 -- 2.54 \\
\multicolumn{5}{c}{C-type stars (C/O\,$>$\,1.1; {\tt Mode = 0,1})$^\ast$}\\ %extracts-c.dat
Age (Gyr)                   & 0.15 & 0.13 -- 0.17 & 0.12 -- 0.18 & 0.12 -- 0.18 \\
$M_{\rm init}$ (M$_\odot$)  & 4.60 & 4.41 -- 4.81 & 4.30 -- 4.98 & 4.26 -- 5.02 \\
$M_{\rm curr}$ (M$_\odot$)  & 1.85 & 1.28 -- 2.48 & 1.03 -- 3.01 & 0.97 -- 3.22 \\
\hline
\multicolumn{5}{p{0.4\textwidth}}{$^\ast$Ranges for these entires are approximate, given the coarse metallicity scaling, and are applicable only within $\sim$100\,pc of the Sun.}\\
\hline
\end{tabular}
\end{table}
% stats.bash extracts-*.dat wjk

\begin{table}
    \centering
    \caption{As Table \ref{tab:typical}, but weighted by mass-loss rate.}
    \label{tab:typicalMLR}
    \begin{tabular}{@{}l@{\quad}c@{\quad}c@{\quad}c@{\quad}c@{}}
    \hline \hline
Parameter                   & Median    & 1$\sigma$ range   & 2$\sigma$ range   & 3$\sigma$ range \\
\hline
\multicolumn{5}{c}{All stars ({\tt Mode = 0-4})}\\ %extracts-all.dat
Age (Gyr)                   & 9.33 & 5.89 -- 11.2 & 2.95 -- 12.6 & 0.57 -- 13.2 \\
$M_{\rm init}$ (M$_\odot$)  & 1.06 & 1.00 -- 1.21 & 0.96 -- 1.51 & 0.94 -- 2.88 \\
$M_{\rm curr}$ (M$_\odot$)  & 0.99 & 0.87 -- 1.14 & 0.74 -- 1.41 & 0.61 -- 2.20 \\
\multicolumn{5}{c}{Mass-losing stars ($\dot{M} > 10^{-7}$\,M$_\odot$\,yr$^{-1}$)}\\ %extracts-masslosing.dat
Age (Gyr)                   & 9.55 & 3.72 -- 11.5 & 0.85 -- 12.9 & 0.28 -- 13.2 \\
$M_{\rm init}$ (M$_\odot$)  & 1.05 & 0.99 -- 1.40 & 0.96 -- 2.53 & 0.94 -- 3.66 \\
$M_{\rm curr}$ (M$_\odot$)  & 0.82 & 0.74 -- 0.99 & 0.63 -- 1.85 & 0.59 -- 2.91 \\
\multicolumn{5}{c}{LPVs ($P > 60$\,days)}\\ %extracts-gt60days.dat
Age (Gyr)                   & 9.55 & 3.39 -- 11.5 & 0.79 -- 12.9 & 0.26 -- 13.2 \\
$M_{\rm init}$ (M$_\odot$)  & 1.05 & 0.99 -- 1.45 & 0.96 -- 2.59 & 0.94 -- 3.74 \\
$M_{\rm curr}$ (M$_\odot$)  & 0.81 & 0.73 -- 1.02 & 0.63 -- 1.89 & 0.59 -- 2.94 \\
\multicolumn{5}{c}{LPVs ($P > 60$\,days) above RGB tip ($L > 2500$\,L$_\odot$)}\\ %extracts-gt60days-trgb.dat
Age (Gyr)                   & 5.37 & 1.74 -- 9.77 & 0.57 -- 11.5 & 0.16 -- 12.9 \\
$M_{\rm init}$ (M$_\odot$)  & 1.25 & 1.05 -- 1.98 & 1.00 -- 2.88 & 0.96 -- 4.49 \\
$M_{\rm curr}$ (M$_\odot$)  & 0.86 & 0.71 -- 1.33 & 0.61 -- 2.17 & 0.59 -- 3.30 \\
\multicolumn{5}{c}{Sequences $B$, $C$ and $C^\prime$ ({\tt Mode = 0,1})}\\ %extracts-bccp.dat
Age (Gyr)                   & 9.77 & 3.47 -- 11.5 & 0.81 -- 12.9 & 0.27 -- 13.2 \\
$M_{\rm init}$ (M$_\odot$)  & 1.05 & 0.99 -- 1.43 & 0.96 -- 2.57 & 0.94 -- 3.72 \\
$M_{\rm curr}$ (M$_\odot$)  & 0.81 & 0.73 -- 1.01 & 0.63 -- 1.79 & 0.59 -- 2.92 \\
\multicolumn{5}{c}{Sequence $C^\prime$ ({\tt Mode = 0})}\\ %extracts-cp.dat
Age (Gyr)                   & 9.77 & 3.47 -- 11.5 & 0.81 -- 12.9 & 0.26 -- 13.2 \\
$M_{\rm init}$ (M$_\odot$)  & 1.05 & 0.99 -- 1.43 & 0.96 -- 2.57 & 0.94 -- 3.74 \\
$M_{\rm curr}$ (M$_\odot$)  & 0.81 & 0.73 -- 1.00 & 0.63 -- 1.80 & 0.59 -- 2.92 \\
\multicolumn{5}{c}{O-rich stars (C/O\,$<$\,1; $L > 2500$\,L$_\odot$)$^\ast$}\\ %extracts-orich.dat
Age (Gyr)                   & 5.50 & 1.95 -- 9.77 & 0.89 -- 11.5 & 0.20 -- 12.9 \\
$M_{\rm init}$ (M$_\odot$)  & 1.23 & 1.04 -- 1.85 & 0.99 -- 2.49 & 0.96 -- 4.13 \\
$M_{\rm curr}$ (M$_\odot$)  & 0.85 & 0.70 -- 1.29 & 0.61 -- 2.17 & 0.59 -- 3.31 \\
\multicolumn{5}{c}{C-rich stars (C/O\,$>$\,1; $L > 2500$\,L$_\odot$)$^\ast$}\\ %extracts-crich.dat
Age (Gyr)                   & 0.73 & 0.56 -- 0.91 & 0.49 -- 1.02 & 0.13 -- 1.02 \\
$M_{\rm init}$ (M$_\odot$)  & 2.67 & 2.48 -- 2.90 & 2.38 -- 3.03 & 2.38 -- 4.89 \\
$M_{\rm curr}$ (M$_\odot$)  & 1.55 & 1.04 -- 2.03 & 0.88 -- 2.34 & 0.81 -- 3.11 \\
\multicolumn{5}{c}{M-type stars (C/O\,$<$\,0.6; {\tt Mode = 0,1})$^\ast$}\\ %extracts-m.dat
Age (Gyr)                   & 9.77 & 4.37 -- 11.8 & 1.70 -- 12.9 & 1.10 -- 13.2 \\
$M_{\rm init}$ (M$_\odot$)  & 1.04 & 0.99 -- 1.33 & 0.96 -- 2.00 & 0.94 -- 2.32 \\
$M_{\rm curr}$ (M$_\odot$)  & 0.81 & 0.73 -- 0.95 & 0.63 -- 1.48 & 0.59 -- 2.19 \\
\multicolumn{5}{c}{S-type stars (0.6\,$<$ C/O\,$<$ 1; {\tt Mode = 0,1})$^\ast$}\\ %extracts-s.dat
Age (Gyr)                   & 1.10 & 0.46 -- 1.26 & 0.25 -- 1.32 & 0.11 -- 1.32 \\
$M_{\rm init}$ (M$_\odot$)  & 2.32 & 2.22 -- 3.10 & 2.19 -- 3.81 & 2.19 -- 5.20 \\
$M_{\rm curr}$ (M$_\odot$)  & 1.65 & 1.04 -- 2.30 & 0.74 -- 2.83 & 0.68 -- 4.11 \\
\multicolumn{5}{c}{SC-type stars (1\,$<$ C/O\,$<$ 1.1; {\tt Mode = 0,1})$^\ast$}\\ %extracts-sc.dat
Age (Gyr)                   & 0.74 & 0.56 -- 0.93 & 0.49 -- 1.02 & 0.14 -- 1.02 \\
$M_{\rm init}$ (M$_\odot$)  & 2.65 & 2.46 -- 2.90 & 2.38 -- 3.03 & 2.38 -- 4.72 \\
$M_{\rm curr}$ (M$_\odot$)  & 1.57 & 1.04 -- 2.04 & 0.88 -- 2.33 & 0.81 -- 3.10 \\
\multicolumn{5}{c}{C-type stars (C/O\,$>$\,1.1; {\tt Mode = 0,1})$^\ast$}\\ %extracts-c.dat
Age (Gyr)                   & 0.15 & 0.13 -- 0.17 & 0.12 -- 0.18 & 0.12 -- 0.18 \\
$M_{\rm init}$ (M$_\odot$)  & 4.60 & 4.41 -- 4.85 & 4.30 -- 5.02 & 4.26 -- 5.02 \\
$M_{\rm curr}$ (M$_\odot$)  & 1.84 & 1.27 -- 2.47 & 1.02 -- 3.00 & 0.97 -- 3.22 \\
\hline
\multicolumn{5}{p{0.4\textwidth}}{$^\ast$Ranges for these entires are approximate, given the coarse metallicity scaling, and are applicable only within $\sim$100\,pc of the Sun.}\\
\hline
\end{tabular}
\end{table}
% stats.bash extracts-*.dat wjk w

\subsection{AGB stars with known properties}
\label{sec:cluster}

\begin{figure}
    \centering
    \includegraphics[height=0.47\textwidth,angle=-90]{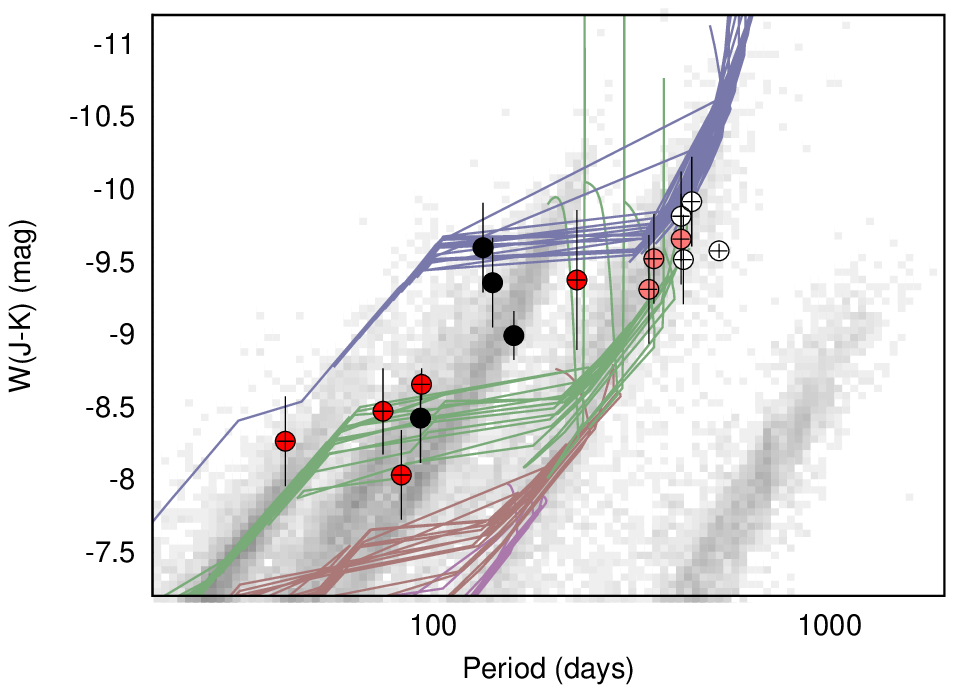}
    \caption{AGB stars in open clusters (large points), with assumed error bars, overlain on the upper portion of the $P-L$ diagram from Figure \ref{fig:pl}. Evolutionary tracks for stars of ages 4.5, 1.6 and 0.28\,Gyr are respectively shown (bottom to top) as red, green and blue lines. Shading on the points represents (from white through black) stars with nominal ages correctly predicted to within the 1$\sigma$, 2$\sigma$, 3$\sigma$ or $>$3$\sigma$ confidence intervals.} 
    \label{fig:plcluster}
\end{figure}

\begin{table*}
    \centering
    \caption{Predicted ages of AGB stars in open clusters, in order of nominal age.}
    \label{tab:cluster}
    \begin{tabular}{@{}l@{\ }l@{\quad}c@{\ }c@{\quad}c@{\ }c@{\quad}c@{\ }c@{\quad}c@{\quad}c@{\quad}c@{\quad}c@{\ }c@{\quad}c@{\ }c@{\quad}c@{\ }c@{\quad}c@{}}
    \hline \hline
Cluster	& 2MASS	& \multicolumn{8}{c}{Inputs from \citep{Marigo22}}	& \multicolumn{7}{c}{Estimated ages and confidence intervals$^2$}	& $n\sigma$\\
Name	& Identifier	& $P$	& $dP$	& $M_J$	& $dM_J$	& $M_{Ks}$	& $dM_{Ks}$	& Age$^2$	& $M_{\rm init}$	& Median	& $-1\sigma$	& $+1\sigma$	& $-2\sigma$	& $+2\sigma$	& $-3\sigma$	& $+3\sigma$	& $^1$\\
\hline
King 11  &	023475728+6835426 &	85.02 &	4.251 &	$-$6.06 &	0.21 &	$-$7.23 &	0.21 &	4.46 &	1.31 &	7.94 &	5.37 &	11.22 &	4.57 &	12.59 &	0.43 &	13.18 &	3 \\
Trumpler 5  &	06363268+0925393 &	432.4 &	21.62 &	$-$4.70 &	0.22 &	$-$7.64 &	0.22 &	4.26 &	1.33 &	2.13 &	0.87 &	4.07 &	0.03 &	6.02 &	0.01 &	7.41 &	2 \\
Pismis 3  &	08310560$-$3838107 &	460 &	23 &	$-$4.57 &	0.21 &	$-$7.74 &	0.21 &	3.16 &	1.47 &	1.77 &	0.58 &	3.71 &	0.03 &	5.37 &	0.01 &	6.60 &	1 \\
BH 55  &	08561346$-$3930429 &	43.28 &	2.164 &	$-$6.14 &	0.21 &	$-$7.40 &	0.21 &	2.29 &	1.65 &	9.55 &	4.57 &	11.48 &	3.31 &	12.88 &	1.02 &	13.18 &	3 \\
Ruprecht 37  &	07494578$-$1715141 &	235.9 &	11.795 &	$-$6.34 &	0.33 &	$-$8.14 &	0.33 &	2.23 &	1.66 &	10.23 &	8.13 &	11.75 &	3.54 &	12.88 &	1.38 &	13.18 &	3 \\
Dias 2  &	06090764+0436414 &	539 &	26.95 &	$-$6.12 &	0.03 &	$-$8.17 &	0.03 &	1.73 &	1.84 &	1.31 &	0.39 &	3.02 &	0.02 &	4.26 &	0.01 &	5.01 &	1 \\
NGC 1798  &	05113360+4740468 &	76.36 &	3.818 &	$-$6.38 &	0.20 &	$-$7.62 &	0.20 &	1.65 &	1.93 &	6.91 &	5.24 &	9.77 &	4.57 &	12.02 &	0.70 &	13.18 &	3 \\
NGC 2660  &	08423384$-$4712252 &	438.2 &	21.91 &	$-$6.65 &	0.21 &	$-$8.35 &	0.21 &	1.62 &	1.94 &	2.09 &	0.83 &	3.98 &	0.03 &	5.88 &	0.01 &	7.24 &	1 \\
NGC 7789  &	05113360+4740468 &	432.5 &	21.625 &	$-$6.61 &	0.21 &	$-$8.51 &	0.21 &	1.54 &	1.97 &	2.13 &	0.83 &	3.98 &	0.03 &	5.88 &	0.01 &	7.41 &	1 \\
Berkeley 9  &	03325580+5244137 &	163.4 &	8.17 &	$-$6.48 &	0.11 &	$-$7.97 &	0.11 &	1.38 &	2.06 &	9.77 &	7.76 &	11.48 &	6.16 &	12.88 &	5.37 &	13.18 &	$\cdots$ \\
Tombaugh 1  &	07001366$-$2033294 &	95.52 &	4.776 &	$-$6.65 &	0.07 &	$-$7.84 &	0.07 &	1.25 &	2.13 &	10.71 &	8.31 &	12.02 &	4.89 &	12.88 &	0.64 &	13.18 &	3 \\
Berkeley 53  &	20560804+5104316 &	358.4 &	17.92 &	$-$6.68 &	0.26 &	$-$8.24 &	0.26 &	0.97 &	2.33 &	3.38 &	1.47 &	5.37 &	0.64 &	7.76 &	0.03 &	10.00 &	2 \\
NGC 2533  &	08070513$-$2947435 &	368.9 &	18.445 &	$-$6.79 &	0.21 &	$-$8.41 &	0.21 &	0.70 &	2.59 &	3.23 &	1.34 &	5.12 &	0.58 &	7.41 &	0.03 &	9.77 &	2 \\
BH 67  &	09264885$-$5114107 &	144.4 &	7.22 &	$-$7.08 &	0.21 &	$-$8.43 &	0.21 &	0.61 &	2.72 &	9.77 &	7.58 &	11.48 &	6.02 &	12.59 &	5.37 &	13.18 &	$\cdots$ \\
JS 1 &	00161695+5958115 &	95.01 &	4.7505 &	$-$6.40 &	0.21 &	$-$7.60 &	0.21 &	0.60 &	2.74 &	10.71 &	8.13 &	12.02 &	4.89 &	12.88 &	0.70 &	13.18 &	$\cdots$ \\
Haffner 14  &	07444463$-$2824077 &	136.7 &	6.835 &	$-$7.17 &	0.21 &	$-$8.61 &	0.21 &	0.28 &	3.54 &	9.77 &	7.76 &	11.48 &	6.02 &	12.59 &	5.37 &	13.18 &	$\cdots$ \\
\hline
\multicolumn{18}{p{0.95\textwidth}}{$^1$Minimum integer number of $\sigma$ at which the value agrees with the nominal age for the cluster in \citet{Marigo22}. Uncertainties in the nominal age are not taken into account, but are expected to a factor of $\sim\pm$74 per cent \citep{Abia25}. $^2$All ages are given in Gyr.}\\
\hline
\end{tabular}
\end{table*}

Testing the ability to derive the masses of Galactic AGB stars from the $P-L$ diagram is difficult, as the sample of stars with known masses is both extremely small, and limited mostly to either stellar populations that don't represent the bulk Galaxy (e.g., globular clusters), or binary stars where interaction may affect evolution.

A few AGB stars are found in open clusters: Table \ref{tab:cluster} lists the 16 stars from \citet{Marigo22} with known periods. We estimate the error in adopted period to be $\sim$5 per cent, while we use the different cases from \citet{Abia25} to determine errors in absolute $J$ and $K_{\rm s}$ magnitude (where none are available for a specific star, the average of 0.218\,mag is used). The latter errors are comparatively large, primarily due to the uncertainties in both distance and the substantial interstellar reddening correction towards some of these clusters.

Table \ref{tab:cluster} also lists the results of our age estimation for these stars. It is notable that only 4 out of 16 of stars (25 per cent) have ages correctly estimated to within 1$\sigma$ (68 per cent confidence interval), while the numbers only rise to 7/16 (44 per cent) for 2$\sigma$ (95 per cent) and 12/16 (75 per cent) for 3$\sigma$ (99.5 per cent).

Three factors likely give rise to this under-performance. Firstly, we have not accounted for errors in the nominal ages listed in \citet{Marigo22}. These are not available for all the clusters in the list, so we cannot fully incorporate them into our comparison. However, the few uncertainties that are available indicate that the cluster age is typically only accurate to $\sim\pm$0.24 dex \citep{Abia25}. Given our typical error in age is $\pm$0.18 dex, these should add considerably to the overall error budget.

Secondly, any uncertainty in period, magnitude or distance will cause a spread in statistical properties, thereby artificially making the properties of the sample more closely resemble the underlying observed AGB mass function (Figure \ref{fig:massfn}; Table \ref{tab:typical}). Since most of these stars are well above the expected median mass of Galactic AGB stars, they are intrinsically rare objects. Any field stars with the same position on the $P-L$ diagram are indeed likely to be closer to the older ages predicted by our $P-L$ method.

Finally, Figure \ref{fig:plcluster} shows these stars on the $P-L$ diagram. Stars on the fundamental mode generally have well-estimated ages, while first-overtone stars have poorly estimated ages. Figure \ref{fig:plcluster} also shows comparison tracks corresponding to the youngest, median and oldest star among the 16. Most stars lie on or above the median track, indicating that the models tend to evolve the star onto the fundamental mode before they are actually observed there. (This may also be an observational problem, if the wrong pulsation mode has been identified as dominant in a multi-mode pulsator.)

Consequently, care should be taken when using this method for individual stars and is better for relative comparisons of groups of stars. This test has shown that the method has high specificity towards younger, more-massive stars, but it has low sensitivity to them.

\subsection{Statistical samples of AGB stars}

To better demonstrate this method's statistical power, we have drawn samples of AGB stars from the literature. To determine individual masses, we convolve the ``All stars'' distributions in Table \ref{tab:typical} / Figure \ref{fig:lifetime} with the pulsation period and absolute magnitudes used by the survey. In doing so, we impose no \emph{a priori} selection on the stars' properties, instead allowing the survey's own selection criteria to determine its probabilistic mass distribution.

Not all stars return non-zero probabilities (some fall outside the $P-L$ relations, e.g., on sequence $D$), so the final number of stars listed in Table \ref{tab:surveys} and the Supplementary Tables contain fewer stars than those with known parameters.

\begin{table}
    \centering
    \caption{Statistical properties of AGB stars from different surveys. No chemical data has been used.}
    \label{tab:surveys}
    \begin{tabular}{@{}l@{\quad}c@{\quad}c@{\quad}c@{\quad}c@{}}
    \hline \hline
Parameter                   & Median    & 1$\sigma$ range   & 2$\sigma$ range   & 3$\sigma$ range \\
\hline
%\multicolumn{5}{c}{\emph{Gaia} long-period variables (2994 stars)}\\
%Age (Gyr)                   & 9.55     & 6.61 -- 11.5    & 2.46 -- 12.9    & 0.23 -- 13.2 \\
%$M_{\rm init}$ (M$_\odot$)  & 1.05     & 0.99 -- 1.17     & 0.96 -- 1.64     & 0.94 -- 3.93 \\
%$M_{\rm curr}$ (M$_\odot$)  & 0.98     & 0.91 -- 1.10     & 0.79 -- 1.49     & 0.68 -- 3.24 \\
\multicolumn{5}{c}{\emph{Gaia} long-period variables $\times$ GCVS (217 stars)}\\
Age (Gyr)                   & 8.91     & 3.80 -- 11.2     & 0.66 -- 12.6     & 0.025 -- 13.2 \\
$M_{\rm init}$ (M$_\odot$)  & 1.07     & 1.00 -- 1.39     & 0.96 -- 2.75     & 0.94 -- 9.96 \\
$M_{\rm curr}$ (M$_\odot$)  & 0.99     & 0.91 -- 1.20     & 0.78 -- 2.27     & 0.69 -- 9.95 \\
\multicolumn{5}{c}{300\,pc sample (356 stars)}\\
Age (Gyr)                   & 8.13     & 3.55 -- 11.0     & 0.38 -- 12.6     & 0.010 -- 13.2 \\
$M_{\rm init}$ (M$_\odot$)  & 1.10     & 1.01 -- 1.41     & 0.96 -- 3.27     & 0.94 -- 17.4 \\
$M_{\rm curr}$ (M$_\odot$)  & 1.04     & 0.94 -- 1.33     & 0.83 -- 3.21     & 0.72 -- 15.0 \\
\multicolumn{5}{c}{NESS sample (464 stars)}\\
Age (Gyr)                   & 4.90     & 0.81 -- 10.5     & 0.019 -- 12.3     & 0.010 -- 13.2 \\
$M_{\rm init}$ (M$_\odot$)  & 1.28     & 1.02 -- 2.57     & 0.97 -- 11.7     & 0.95 -- 17.4 \\
$M_{\rm curr}$ (M$_\odot$)  & 1.08     & 0.93 -- 2.03     & 0.80 -- 11.7     & 0.70 -- 15.0 \\
\multicolumn{5}{c}{DEATHSTAR sample (143 stars)}\\
Age (Gyr)                   & 3.98     & 1.02 -- 10.24     & 0.030 -- 12.3     & 0.011 -- 13.2 \\
$M_{\rm init}$ (M$_\odot$)  & 1.37     & 1.03 -- 2.38     & 0.97 -- 9.19     & 0.95 -- 17.0 \\
$M_{\rm curr}$ (M$_\odot$)  & 1.07     & 0.92 -- 1.95     & 0.81 -- 9.18     & 0.70 -- 14.8 \\
\multicolumn{5}{c}{ATOMIUM sample (16 stars)}\\
Age (Gyr)                   & 2.34     & 0.34 -- 8.51     & 0.020 -- 12.0     & 0.011 -- 13.2 \\
$M_{\rm init}$ (M$_\odot$)  & 1.69     & 1.08 -- 3.43     & 0.98 -- 11.1     & 0.95 -- 17.0 \\
$M_{\rm curr}$ (M$_\odot$)  & 1.32     & 0.95 -- 2.81     & 0.84 -- 11.1     & 0.72 -- 14.9 \\
\hline
\end{tabular}
\end{table}

\begin{table}
    \centering
    \caption{Statistical properties of NESS survey tiers.}
    \label{tab:NESS}
    \begin{tabular}{@{}l@{\quad}c@{\quad}c@{\quad}c@{\quad}c@{}}
    \hline \hline
Parameter                   & Median    & 1$\sigma$ range   & 2$\sigma$ range   & 3$\sigma$ range \\
\hline
\multicolumn{5}{c}{Tier 4 (``extreme'')}\\
Age (Gyr)                  & 0.54      & 0.011 -- 2.14      & 0.010 -- 4.37     & 0.010 -- 6.46 \\
$M_{\rm init}$ (M$_\odot$) & 2.94      & 1.78 -- 16.4      & 1.33 -- 17.4      & 1.18 -- $\infty$ \\
$M_{\rm curr}$ (M$_\odot$) & 2.21      & 1.26 -- 14.4      & 0.98 -- 15.0      & 0.87 -- $\infty$ \\
\multicolumn{5}{c}{Tier 3 (``high'')}\\
Age (Gyr)                  & 2.14      & 0.40 -- 7.95      & 0.017 -- 12.0     & 0.010 -- 13.2 \\
$M_{\rm init}$ (M$_\odot$) & 1.78      & 1.11 -- 3.25      & 0.98 -- 12.6      & 0.95 -- 17.3 \\
$M_{\rm curr}$ (M$_\odot$) & 1.32      & 0.95 -- 2.76      & 0.82 -- 12.2      & 0.71 -- 15.0 \\
\multicolumn{5}{c}{Tier 2 (``intermediate'')}\\
Age (Gyr)                  & 8.13      & 2.57 -- 11.2      & 0.038 -- 12.6      & 0.010 -- 13.2 \\
$M_{\rm init}$ (M$_\odot$) & 1.10      & 1.00 -- 1.63      & 0.97 -- 8.24      & 0.94 -- 17.4 \\
$M_{\rm curr}$ (M$_\odot$) & 0.99      & 0.91 -- 1.28      & 0.78 -- 8.23      & 0.69 -- 15.0 \\
\multicolumn{5}{c}{Tier 1 (``low'')}\\
Age (Gyr)                  & 8.51      & 4.57 -- 11.0      & 0.81 -- 12.6      & 0.021 -- 13.2 \\
$M_{\rm init}$ (M$_\odot$) & 1.09      & 1.01 -- 1.30      & 0.96 -- 2.57      & 0.94 -- 10.8 \\
$M_{\rm curr}$ (M$_\odot$) & 1.04      & 0.95 -- 1.26      & 0.87 -- 2.04      & 0.74 -- 10.8 \\
\multicolumn{5}{c}{Tier 0 (``very low'')}\\
Age (Gyr)                  & 9.33      & 4.79 -- 11.5      & 0.54 -- 12.9      & 0.33 -- 13.2 \\
$M_{\rm init}$ (M$_\odot$) & 1.06      & 1.00 -- 1.29      & 0.96 -- 2.94      & 0.94 -- 3.45 \\
$M_{\rm curr}$ (M$_\odot$) & 1.00      & 0.91 -- 1.26      & 0.78 -- 2.94      & 0.69 -- 3.45 \\
\hline
\end{tabular}
\end{table}

\subsubsection{\emph{Gaia} long-period variables}

\emph{Gaia}'s 2023 Focussed Product Release included a set of pulsation frequencies and radial-velocity data for Galactic long-period variables \citep[LPV;][VizieR table {\tt I/361/vlpv}]{GaiaLPV}. This was cross-matched against parallax data from \emph{Gaia} Data Release 3 \citep[Vizier table {\tt I/355/gaiadr3}]{GaiaDR3} and the Two-Micron All Sky Survey \citep[2MASS;][Vizier table {\tt II/246/out}]{Cutri2003} using a 1$^{\prime\prime}$ search radius.

From this, a list of high-quality sources were chosen, with 2MASS $J$ and $K_{\rm s}$ magnitudes accurate to within 0.3 mag, and parallaxes accurate to better than 10 per cent, giving 3304 stars. This latter category ensures that the uncertainty in distance does not substantially affect the fit, and so that symmetric errors can be used in absolute magnitude. Unfortunately, the errors in frequency in this dataset remain too large ($\delta f \approx f$), so many stars are localised to the short-period side of the diagram, resulting in unreasonably high masses and young ages.

% Since we have selected LPV stars, we expect these properties to most-closely resemble the ``Sequence $B$, $C$ and $C^\prime$ stars'' cut. Indeed, the median observed age and initial mass are only slightly younger and higher than the expected ones. However, the current mass is significantly higher than expected, suggesting that the large errors in pulsation frequency have shifted the distribution of all three parameters closer to the ``All Stars'' distribution, and that the bias may be larger.

To reduce these errors, the \emph{Gaia} LPV dataset was further cross-matched this dataset against the (generally more accurate) periods of the General Catalogue of Variable Stars \citep[GCVS;][]{Samus17} using a 2$^{\prime\prime}$ radius. Only stars with measured periods were retained, and a period uncertainty of 5 per cent was assumed (the GCVS does not provide a period error directly). Applying the same cuts, this retains only 221 stars and yields the second set of values in Table \ref{tab:surveys}. 

The resulting median properties and statistical ranges of \emph{Gaia} LPV stars can be seen in Table \ref{tab:surveys}. We expect the properties to broadly match the ``LPVs'' and ``Sequences $B$, $C$ and $C^\prime$'' categories from Table \ref{tab:typical} which, by and large, they do. We do see a very slight preference for younger ages and higher masses, which has several possible explanations. This could include a skew towards the ``All stars'' parameters due to finite uncertainties in the data, but also a potential \citet{Malmquist22} bias for detecting brighter, higher-mass stars and a potential preference among historical observers for recording higher-amplitude Mira variables.

\subsubsection{NESS}

NESS (the Nearby Evolved Stars Survey; \citealt{Scicluna21}) is a volume-limited sample with a wedding-cake design, with five tiers limited by both distance and mass-loss rate. The survey includes a catalogue of 781 AGB stars \citep{McDonald2025}, from which a sub-sample of 398 were chosen that have known periods (again, a 5 per cent error was assumed), known $J$ and $K_{\rm s}$ magnitudes from 2MASS and distances accurate to within 30 per cent. Where 2MASS magnitudes were not available, $F_{1.25}$ and $F_{2.2}$ magnitudes were substituted from the Diffuse Infrared Background Experiment on the \emph{Cosmic Background Explorer} (\emph{COBE}; \citealt{Smith2004}; to a total of 472 stars): these filters have similar central wavelengths to the 2MASS filters but are broader than them.

\citet{McDonald2025} also includes a partially overlapping sample of 507 evolved (both red giant branch, RGB, and AGB) stars within 300\,pc of the Sun. Within this sample, there are 384 stars that meet the same criteria as used for the NESS main-catalogue stars above. We expect the full set of 507 stars to follow the ``All Stars'' distribution, but the subsample of 384 should more closely follow the ``LPVs'' cut. This sample should be the closest representation to the adopted SFH (which is based on stars within 100\,pc of the Sun).

The statistical distribution of age and mass for these samples are listed in Table \ref{tab:surveys}. The 300\,pc sample is again slightly younger than the expected distribution for a random sample of AGB in the Galaxy, and more closely resembles the ``All stars'' distribution, though with slightly broader ranges, particularly in the young, high-mass tails. Similar possible explanations apply as for the \emph{Gaia} LPV sample.

The NESS survey proper has different expectations, which vary by tier due to the different mass-loss rates (and, to a lesser extent, the volumes sampled). Individual tiers can be relatively small. However, distinct differences are still clearly seen between the tiers (see Table \ref{tab:NESS}). The very-low-mass-loss-rate, low-mass-loss rate and intermediate-mass-loss-rate Tiers 0, 1 and 2 all match expectations for large populations of stars (Table \ref{tab:typical}), and have similar properties to both the \emph{Gaia} LPV and 300\,pc sample (Table \ref{tab:surveys}). The age of the star decreases slightly (and the mass increases correspondingly) between tiers, as expected for the progressively higher mass-loss-rate cutoffs. However, it is only in the high- and extreme-mass-loss-rate Tiers 3 and 4 where the natural bias towards younger and higher-mass populations really appears, with the ages of sources becoming factors of several lower and older, low-mass stars disappearing from the 1$\sigma$ and (for Tier 4) 2$\sigma$ ranges.

Overall, this biases the NESS survey towards younger stars. This was partly intentional, as the survey is meant to sample stars by their contribution to the ISM, rather than by number. The counterparts of the NESS survey's``extreme'' Tier 4 AGB stars are the dominant dust-producers in other (metal-poor) Local Group galaxies \citep[e.g.][]{Boyer15}. However, it is not clear that this holds true in our Galaxy: our weighted models indicate that the median contributor to the ISM is a star losing mass at only $2-3 \times 10^{-6}$\,M$_\odot$\,yr$^{-1}$ (corresponding to a dust-production rate of $\sim1-2 \times 10^{-8}$\,M$_\odot$\,yr$^{-1}$, within the NESS ``high'' Tier 3). With an average initial mass of 1.28\,M$_\odot$ compared to the average mass-losing star of 1.22\,M$_\odot$ (Section \ref{sec:results}), the NESS survey therefore seems to have broadly achieved its goal but slightly over-estimated the contribution from these more-extreme sources.

\subsubsection{DEATHSTAR}

DEATHSTAR (``Determining accurate mass-loss rates for thermally pulsing AGB stars''; \citet{Ramstedt2020}) is a large programme using the Atacama Large Millimetre Array's Compact Array (ALMA/ACA) to map known AGB stars. These stars have been taken from existing samples, therefore broadly reflect the properties of AGB stars studied in preceding literature. Its selection partly balances M-type, S-type and C-type stars, so as to sample the differences in mass loss and chemical enrichment in each type.

Distances to these stars were published in \citet{Andriantsaralaza2022}, which is here cross-matched against 2MASS and GCVS using a 2$^{\prime\prime}$ radius, using the same restrictions in magnitude and distance errors as previously. Again, a 5 per cent error on period was assumed. Upper and lower distance uncertainties were averaged which, when converted to magnitudes, approximately accounts for the asymmetric uncertainties provided $\delta d \ll d$. This results in a sub-sample of 157 of the 201 stars.

The statistical properties of the DEATHSTAR sample are shown in Table \ref{tab:surveys}. The medians and ranges lie between the NESS ``intermediate'' and ``high'' tiers. The DEATHSTAR survey therefore almost entirely samples the younger, higher-mass half of AGB stars and is strongly biased towards younger, higher-mass and more extreme stars overall. This is partly a result of the historical bias in the literature towards higher-mass-loss-rate stars, but the bias is probably dominated by the intentional over-selection in DEATHSTAR of carbon and S-type stars: note that chemical information has not been added into these mass estimates: when combined with the aforementioned bias towards median values incurred by uncertainties in $P$ and $L$, the true bias of DEATHSTAR towards younger, higher-mass stars is likely more significant than these statistics suggest.

\subsubsection{ATOMIUM}

ATOMIUM (``ALMA tracing the origins of molecules in dust-forming, oxygen-rich, M-type stars''; \citet{Decin2020,Gottlieb2022}) observed a small sample of nearby AGB stars at high angular resolution with ALMA. The sample was constructed to be representative of different classes, but not necessarily of AGB stars in general. Again, a 5 per cent error on period was assumed. The error in distance from \citet{Andriantsaralaza2022} was assumed if available, otherwise a 10 per cent distance error was assumed. These data were cross-matched against 2MASS using a 5$^{\prime\prime}$ search radius.

Table \ref{tab:surveys} shows that the ATOMIUM stars are strongly biased to stars of very young ages (largely due to the inclusion of red supergiants in the sample). However, due to the small sample size, the results for ATOMIUM should be taken as indicative only.

\section{Discussion}
\label{sec:disc}

\subsection{Implications for AGB-star samples}

The overall properties of the nominally unbiased samples in Table \ref{tab:surveys} (\emph{Gaia} LPV and 300\,pc samples) broadly reflect the expected averages in Table \ref{tab:typical} or even Table \ref{tab:typicalMLR}, demonstrating that the inclusion of pulsation in the Padova models is now at a level suitable enough for measurement of mass for statistical samples of AGB stars. There is a very slight bias in the observations towards younger ages and higher masses than predicted, which could have a number of observational origins, but may also be due to slight inaccuracies in our chosen SFH and its application to these different parts of our Galaxy, or it may reflect comparatively small errors in the notoriously difficult practice of modelling AGB stars in the stellar evolution models.

However, much larger differences exists when modelled expectations are compared against survey samples like NESS, DEATHSTAR and ATOMIUM. While these surveys do not purport to be numerically representative of AGB stars or their mass-loss properties, it is apparent that they all have an overall bias towards brighter, younger and more-massive stars, which typically have higher mass-loss rates. The factors behind this are complex, but likely stem from two sources: (1) the over-representation of chemically peculiar stars in the literature, by the simple virtue that their peculiarity is scientifically interesting; and (2) the over-representation of stars with high mass-loss rates in the literature.

Care should therefore be taken in applying the results of samples like these surveys to AGB stars as a whole. In the case of NESS in particular, this can be accomplished by examining the individual tiers separately and weighting appropriately. For DEATHSTAR and ATOMIUM, it can be accomplished by appropriate weighting of individual stars.

\subsection{Applicability of the method}

Experiments with individual stars of high mass (Section \ref{sec:cluster}) show that the method does not robustly identify ages of stars that lie a long way from the statistical mean, likely due to both uncertainties in period and absolute magnitude, and to inaccuracies in the models and their associated lack of a one-to-one mapping of age to position in the $P-L$ diagram. Particular case should be taken for overtone pulsators, which may be under-represented in the models.

Based on the above results for the 300\,pc sample (likely the most representative of local AGB stars), the above approach using theoretical $P-L$ diagrams should allow ages for individual AGB stars to be derived to a typical accuracy (median 68.3 per cent confidence interval) of $^{+29}_{-35}$ per cent, initial masses to $^{+14}_{-7}$ per cent and current masses to $^{+17}_{-11}$ per cent. The typical offset from the expected distribution in each of these parameters is, respectively, $-15$, $+7$ and $+27$ per cent, meaning that these systematic errors may dominate (both statistical and systematic errors are smaller if the \emph{Gaia} LPV sample is chosen instead). If these represent meaningful errors for this method, then this offers a very valuable method of assessing initial mass for Galactic AGB stars and improves on methods for assessing age and current mass of AGB stars, and especially allows comparative estimates between samples that avoid these systematic biases.
%awk '$4+0>0 {print $6-$5,$5-$4}' 300pc.ages | sort -nk1 | awk 'NR==179 {print 10**$1-1}'
%awk '$4+0>0 {print $6-$5,$5-$4}' 300pc.ages | sort -nk2 | awk 'NR==179 {print 1-10**-$2}'
%awk '$4+0>0 {print $6/$5,$4/$5}' 300pc.minit | sort -nk1 | awk 'NR==179 {print $1}'
%awk '$4+0>0 {print $6/$5,$4/$5}' 300pc.minit | sort -nk2 | awk 'NR==179 {print $2}'
%awk '$4+0>0 {print $6/$5,$4/$5}' 300pc.mcurr | sort -nk1 | awk 'NR==179 {print $1}'
%awk '$4+0>0 {print $6/$5,$4/$5}' 300pc.mcurr | sort -nk2 | awk 'NR==179 {print $2}'

A substantial contributor to lowering these uncertainty ranges is the relatively peaked SFH, coupled with the IMF, which makes near-solar-mass stars much more common. For higher-mass stars, these confidence intervals increase significantly, with the much larger upper limits deriving from the longer but poorly populated tail of higher-mass stars. When applying this method to galaxies with younger or broader SFHs, we can therefore expect the confidence intervals to be larger than those described in the previous paragraph.
%awk '$2+0>0 {print $6-$5,$5-$4}' ness_t3.ages ness_t4.ages | sort -nk1 | awk 'NR==179 {print 10**$1-1}'
%awk '$2+0>0 {print $6-$5,$5-$4}' ness_t3.ages ness_t4.ages | sort -nk2 | awk 'NR==179 {print 1-10**-$2}'
%awk '$2+0>0 {print $6/$5,$4/$5}' ness_t3.minit ness_t4.minit | sort -nk1 | awk 'NR==179 {print $1}'
%awk '$2+0>0 {print $6/$5,$4/$5}' ness_t3.minit ness_t4.minit | sort -nk2 | awk 'NR==179 {print $2}'
%awk '$2+0>0 {print $6/$5,$4/$5}' ness_t3.mcurr ness_t4.mcurr | sort -nk1 | awk 'NR==179 {print $1}'
%awk '$2+0>0 {print $6/$5,$4/$5}' ness_t3.mcurr ness_t4.mcurr | sort -nk2 | awk 'NR==179 {print $2}'

It is worth emphasising that these systematic errors apply only when the SFH is correctly matched to the observed population. The SFH incorporated into this method, and into Tables \ref{tab:typical} and \ref{tab:typicalMLR} in particular, is valid for stars near the Sun. Applying this to even wider samples of Galactic AGB stars from \emph{Gaia} \citep[cf.][]{Abia22,Mori25} is likely to result in systematic biases due to the much larger fraction of low-mass, metal-poor stars from the Galactic halo incorporated in such surveys. This is especially important if any use of chemical information is included, due to the wider range of masses that form chemically enhanced (S, SC, C) stars at these low metallicities: e.g., the mean mass of S-type stars of $\sim$1.3\,M$_\odot$ in \citet{Mori25} is at odds with 2.63\,M$_\odot$ listed in Table \ref{tab:typical} primarily as a result of this difference, and the much larger number of S-type stars that occur at lower metallicities (which is strongly enhanced by the extreme slope of the current mass function of AGB stars).

This is therefore not a high-accuracy method, but one that is better than other widely applicable tracers of AGB star ages and masses, and one that can be used to great effect in comparing samples, and assesses how well they represent the overall distribution of AGB stars in a galaxy.

\section{Conclusions}
\label{sec:conc}

Evolutionary models that include AGB star pulsations are now sufficiently accurate to derive the properties of both statistical ensembles of AGB stars and, with high specificity but low sensitivity, high-mass outliers from the typical distribution. We demonstrate that a star's position on the $P-L$ diagram can recover its age, initial mass and current mass to within a reasonably small fraction of the true value. These modelled masses and ages can potentially inform currently unknown issues, such as the mass range of stars undergoing hot bottom burning and the efficiency of third dredge-up at different masses. However, because these factors remain poorly determined, we discourage the use of chemical class as Bayesian evidence in the calculation of stellar mass and age using this $P-L$ method.

Based on our assumed IMF and SFH, we can derive the following expectations for Milky Way AGB stars.
\begin{itemize}
    \item The typical star leaving the AGB in the Milky Way had an initial mass of $\approx$1.11\,M$_\odot$. Roughly 95 per cent of today's Galactic AGB stars had initial masses below 2\,M$_\odot$.
    % awk '{t+=$2; print $1,t/s}' s=`awk '{s+=$2} END {print s}' cmf.dat` cmf.dat | more
    \item Weighting stars by their time-integrated mass loss, the median initial mass of AGB stars enriching the Milky Way ISM is $\approx$1.22\,M$_\odot$. Roughly 95 per cent of mass loss from AGB stars comes from stars with initial masses below $\sim$2.8\,M$_\odot$.
    % awk '$1<1 {mf=0.555} $1>=1 && $1<2.5 {mf=0.086*$1+0.469} $1>=2.5 && $1<3.4 {mf=0.1*$1+0.4} $1>=3.4 && $1<5.03 {mf=0.06*$1+0.57} $1>=5.03 {mf=0.17*$1+0.04} {dm=$1-mf; contrib=$2*dm; print $1,contrib}' cmf.dat > foo; awk '{t+=$2; print $1,t/s}' s=`awk '{s+=$2} END {print s}' foo` foo | more
\end{itemize}

Selection criteria of AGB stars can strongly affect whether a sample of AGB stars reflects these distributions. Generally, such selections will lead to distributions of higher mass.
\begin{itemize}
    \item Selection based on pulsation properties appears to have relatively little effect, though this is likely to break down when extreme criteria are used (e.g., if only Miras are selected).
    \item Selection based on presence of circumstellar dust likely has little effect unless only extreme dust producers with substantial circumstellar reddening ($\dot{M} \gg 10^{-7}$\,M$_\odot$\,yr$^{-1}$) are selected. The lack of accurate modelling makes mass-loss rate an unreliable indicator of stellar mass.
    \item Luminosity has a strong effect on the observed AGB mass function, including in the case where luminosities are restricted to stars above the RGB tip.
    \item Chemical type naturally has the strongest effect on the selection of stars of particular masses. However, it is notable that selections of chemically self-enriched stars (S-, SC- and C-type stars) may represent a completely different mass and age range to normal, oxygen-rich stars, which should be taken into account when deriving C/M ratios. The lack of accurate calibration of chemical enrichment makes chemical type a poor indicator of stellar mass at present.
\end{itemize}

Samples of Milky Way AGB stars have differing biases.
\begin{itemize}
    \item Deep optical surveys for variable stars (\emph{Gaia}) and volume-limited surveys of nearby AGB stars (the NESS 300\,pc sample) have mass distributions that largely follow the above expectations. There is a very slight bias towards younger ages and higher masses, which can be explained by any one of several known effects.
    \item Intentional samples of AGB stars, such as the NESS, DEATHSTAR and ATOMIUM surveys are all biased towards stars of considerably higher masses. While there are a number of (often intentional) choices that contribute to this, it is suggested that this bias may come from both historic interest in chemically peculiar stars in the literature and a potentially mistaken understanding that contributions to the Milky Way ISM are dominated by ``extreme'' mass-loss-rate stars. This needs considered when applying these samples to AGB stars in general.
\end{itemize}

\section*{Acknowledgements}

This research has made use of the VizieR catalogue access tool, CDS, Strasbourg, France \citep{10.26093/cds/vizier}. The original description  of the VizieR service was published in \citet{vizier2000}. This research made use of the cross-match service provided by CDS, Strasbourg.

The author acknowledges funding from OSCARS. The OSCARS project has received funding from the European Commission’s Horizon Europe Research and Innovation programme under grant agreement No. 101129751

%%%%%%%%%%%%%%%%%%%%%%%%%%%%%%%%%%%%%%%%%%%%%%%%%%
\section*{Data Availability}

Electronic tables with properties of individual stars, and the scripts used to generate masses from Padova isochrone data, are provided as Supplementary Material to this paper.



\bibliographystyle{mnras}
\bibliography{biblio} % if your bibtex file is called example.bib

%%%%%%%%%%%%%%%%%%%%%%%%%%%%%%%%%%%%%%%%%%%%%%%%%%

% Don't change these lines
\bsp	% typesetting comment
\label{lastpage}
\end{document}